\documentclass[aps,prb,twocolumn,groupedaddress]{revtex4}

\usepackage{graphicx,bm}

\newcommand{\Epara }{$E{\parallel}c$ }
\newcommand{\Eperp  }{$E{\perp}c$ }
\newcommand{\yxxy }{$y$($xx$)$\bar{y}$ }

\begin{document}


\title{Doping variation of anisotropic charge and orbital dynamics 
in Y$_{1-x}$Ca$_x$VO$_3$ : Comparison with La$_{1-x}$Sr$_x$VO$_3$}

\author{J. Fujioka$^{1}$, S. Miyasaka$^{2}$, and Y. Tokura$^{1,3,4}$}
\affiliation{$^{1}$Department of Applied Physics, University of Tokyo, Tokyo 113-8656, Japan}
\affiliation{$^{2}$Department of Physics, University of Osaka, Osaka 560-0043, Japan}

\affiliation{$^{3}$ERATO-Multiferroics Project, Japan Science and Technology Agency, 
Tsukuba, 305-8562, Japan}
\affiliation{$^{4}$Cross-Correlation Materials Research Program (CMRG), RIKEN, 
Wako 351-0198, Japan}

\date{\today}
\begin{abstract}

The doping variation of charge and orbital dynamics 
in perovskite-type Y$_{1-x}$Ca$_x$VO$_3$ ($0\le x\le 0.1$) is investigated 
by measurements of the optical conductivity and Raman scattering spectra 
in comparison with the larger-bandwidth system La$_{1-x}$Sr$_x$VO$_3$. 
We also take into consideration 
the magnitude of the GdFeO$_3$-type orthorhombic lattice distortion, 
which is large and small in Y$_{1-x}$Ca$_x$VO$_3$ and La$_{1-x}$Sr$_x$VO$_3$, 
respectively, and discuss its effect on the evolution of charge dynamics. 
The optical conductivity spectra show that 
the doped hole is well localized and forms the small polaron like state. 
The hole dynamics in Y$_{1-x}$Ca$_x$VO$_3$ is nearly isotropic 
up to the doping level of the orbital order-disorder transition, 
while that in La$_{1-x}$Sr$_x$VO$_3$ is anisotropic 
in the lightly doped region due to 
the one-dimensional orbital exchange interaction. 
The possible origin of the difference in the hole dynamics 
is discussed in terms of the local lattice distortion, which is induced 
by the formation of the small polaron like state 
and becomes more significant for the reduced one-electron bandwidth. 
In addition, 
the optical Mott-gap excitation in the nominally 
$C$-type spin and $G$-type orbital ordered phase is distinct 
from that for La$_{1-x}$Sr$_x$VO$_3$ 
in its intensity and spectral shape. 
This suggests that the orthorhombic lattice distortion 
enhances the modification of the spin and orbital ordering from 
the pure $C$-type and $G$-type, respectively. 
The systematic study of Raman scattering spectra has shown that 
the dynamic $G$-type spin and $C$-type orbital correlation 
subsists at low temperatures 
in the doping induced phase of the nominally $C$-type SO and $G$-type OO. 

\end{abstract}
\pacs{71.30.+h, 72.15.-v, 78.20.-e, 75.30.Et}


\maketitle


\section{\label{sec1}Introduction}

Recent investigations on the 3$d$ transition-metal oxides have revealed 
the important interplay among charge, spin, and orbital degrees of freedom 
in the dramatic change of 
the electronic and magnetic structures, such as  
superconductivity, colossal magnetoresistance, metal-insulator transition, 
and charge-orbital ordering phenomena.\cite{imada_rmp} 
One prototypical way to induce these phenomena is the control of 
the band-filling. 
This may produce the critical state 
where several electronic phases compete with each other 
via the modulation of Coulombic correlation, spin and orbital 
exchange interaction, electron-phonon interaction, and so on. 
For example, the colossal magnetoresistance is induced 
by the competition between the ferromagnetic metallic phase and 
the charge-orbital ordered insulating one.\cite{science,Tokura} 
Recent extensive studies on colossal 
magnetoresistance (CMR) in the perovskite-type manganites have revealed that 
the orbital degree of freedom plays key roles for the versatile magnetic 
phases and charge transport phenomena. 
One prominent feature of the CMR manganites related with the orbital 
degree of freedom is the Jahn-Teller interaction, 
{\it i.e.} the strong coupling of the orbital state 
with the local lattice distortion to lift the orbital degeneracy. 
For example, the transition temperature of 
e$_g$ orbital ordering (OO) in LaMnO$_3$ is as high as 780K, 
as identified as the 
crystal structure transition. 
On the contrary, the spin ordering (SO) occurs at a much lower temperature 
around 140K. \cite{YMurakami_RXS} 
The well separated transition temperatures 
of OO and SO remain to be observed 
even in the hole doped system La$_{1-x}$Sr$_x$MnO$_3$
until the insulator-metal transition. \cite{VAIvanshin2000} 
In other words, 
the energy scale of the OO coupled with lattice distortion 
is much larger than that of SO and 
hence the role of the spin-orbital coupling or quantum nature of the orbital 
degree of freedom is not visible in the system. 

On the other hand, in the perovskite vanadium oxide $R$VO$_3$ 
($R$=rare earth element), which is known as a prototypical 
Mott-Hubbard insulator, the $t_{2g}$ orbital is active and the energy scale 
of the Jahn-Teller interaction is comparable to that of the spin and orbital 
exchange interaction or the on-site spin-orbit interaction. 
This enables 
us to observe the spin-orbital coupled quantum phenomena. 
\cite{kha1, Sirker2003, Horsch2003, Ulrich} 
One prototypical 
example is the one-dimensional confinement of electrons due to the anisitropic 
spin-orbital exchange interaction 
in LaVO$_3$. \cite{kha1, motome} 
In LaVO$_3$, the two valence 
electrons occupy the $3d$ $t_{2g}$ orbitals, 
forming the $S=1$ spin state due to the Hund's-rule coupling. 
The orthorhombic lattice distortion splits the triply degenerate 
$t_{2g}$ levels into the lower-lying $d_{xy}$ state and 
higher-lying doubly degenerate 
$d_{yz}$ and $d_{zx}$ ones.\cite{Mizokawa1996,Sawada1996} 
Thus, one electron occupies the $d_{xy}$ orbital and another 
does either $d_{yz}$ or $d_{zx}$ one. 
With the decrease of temperature, 
the $C$-type SO takes place at $T_{SO1}=143K$ and subsequently 
the $G$-type OO does at $T_{OO1}=141K$. 
Here, the $C$-type spin and $G$-type orbital order means that 
the spins align ferromagnetically along the $c$-axis 
but are alternate in the $ab$-plane, 
while the $d_{zx}$ and $d_{yz}$ orbitals are alternate 
both along the $c$-axis and in the $ab$-plane, 
as schematically shown in Fig. 1.\cite{Zubkov, Bordet, miya2003, Ren2003} 
In the $C$-type spin and $G$-type orbital ordered phase, 
the orbital exchange interaction is quasi-one-dimensional 
due to the interference among the exchange processes, 
leading to the anisotropic feature of the optical 
Mott-gap excitation.\cite{motome, kha_opt, miya_opt} 

By partially replacing the nominally trivalent La ion 
with the divalent Sr one, {\it i.e.} hole doping, 
the filling-controlled insulator-metal transition is achieved 
at the critical doping level of $x_c$=0.176. \cite{Inaba, miya2000} 
The spin-orbital phase diagram in La$_{1-x}$Sr$_x$VO$_3$ is shown 
in Fig. 1 (a). 
\begin{figure}[htbp!]
 \includegraphics[width=3.375in,keepaspectratio=true]{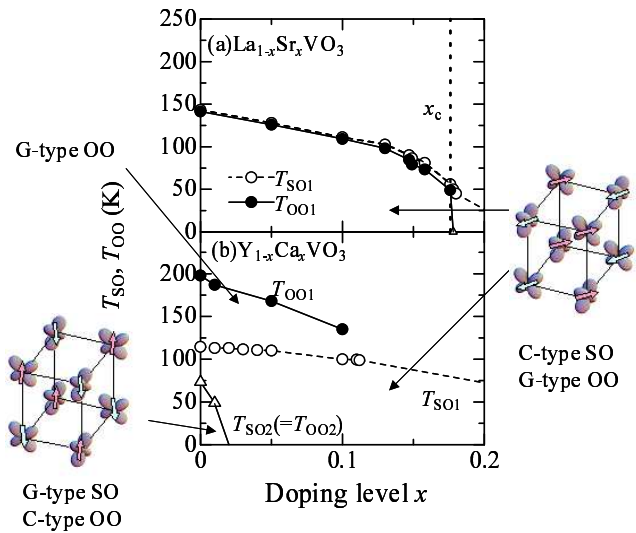}
 \caption{(Color online) The spin and orbital phase diagram of 
(a) La$_{1-x}$Sr$_x$VO$_3$ and (b) Y$_{1-x}$Ca$_x$VO$_3$ plotted against 
the doping level $x$ as reproduced from Refs \cite{miya2000, JF2005}. 
The closed circles with a solid line, open circles with a dashed line, 
and open triangles with a solid line indicate 
the transition temperature of the $G$-type orbital ordering (OO) ($T_{OO1}$), 
the $C$-type spin ordering (SO) ($T_{SO1}$), 
and the $G$-type SO and $C$-type OO ($T_{SO2}=T_{OO2}$), respectively. 
The schematic view shows the pattern of SO and OO. 
The arrows and lobes indicate spins, and $d_{zx}$ or $d_{yz}$ orbitals, 
respectively. 
The vertical dotted line in (a) indicates the critical doping level 
for the insulator-metal transition in La$_{1-x}$Sr$_x$VO$_3$, 
while that for Y$_{1-x}$Ca$_x$VO$_3$ positions around $x$=0.5.}
 \label{fig1}
 \end{figure}
With the increase of the doping level $x$, 
$T_{SO1}$ and $T_{OO1}$ decrease monotonously and 
the $G$-type OO disappears at $x_c$, 
while the $C$-type SO subsists for $x>x_c$. 
Recently, we have investigated the doping variation of 
the electronic structure in the course of 
the insulator-metal transition by measurements of the optical 
conductivity spectra.\cite{JF2006} 
We observed that (1) the hole dynamics is anisotropic 
in the lightly doped region, reflecting the one-dimensional 
orbital exchange interaction, and 
(2) the spin and orbital fluctuation is enhanced 
in the vicinity of the insulator-metal transition, 
leading to the suppression of the one-dimensional electronic structure. 

In another prototypical Mott insulator YVO$_3$ 
with a larger GdFeO$_3$-type orthorhombic lattice distortion 
and hence a smaller 
one-electron bandiwidth, 
the $G$-type OO appears at a much higher temperature 
($T_{OO1}$=200K) than the $C$-type SO ($T_{SO1}$=115K).
\cite{Blake, Noguchi, miya2003} 
Furthermore, 
the subsequent spin-orbital phase transition 
into the $G$-type spin and $C$-type orbital ordered phase 
is observed at $T_{SO2}$(=$T_{OO2}$)=77K. \cite{Kawano} 
Recently, we found that this 
second spin-orbital phase transition is controllable 
not only by changing temperature, but also by the slight variation of 
the band-filling, {\it i.e.} hole doping.\cite{JF2005, Ishihara2005} 
With the increase of $x$ in Y$_{1-x}$Ca$_x$VO$_3$, 
the $G$-type spin and $C$-type orbital ordered 
phase is immediately replaced with the $C$-type spin and $G$-type orbtial 
ordered phase at $x=0.02$, as shown in Fig.1 (b). 
Subsequently the paramangtic 
and $G$-type orbital ordered phase disappears 
around $x_o=0.10$. 
It is anticipated that 
the Ca substitution arouses the quenched disorder in the lattice sector, 
which acts as the random potential for the OO and hence destabilizes the OO. 
This is one possible scenario to explain 
the fact that the critical doping level for the melting 
of the $G$-type OO is smaller in Y$_{1-x}$Ca$_x$VO$_3$ ($x_o$=0.10) 
than in La$_{1-x}$Sr$_x$VO$_3$ ($x$=$x_c$=0.176), 
since the quenched disorder in the lattice sector 
originating from the difference of the ionic radii between $R$ and $A$ ion 
is larger in Y$_{1-x}$Ca$_x$VO$_3$ than in La$_{1-x}$Sr$_x$VO$_3$. 
In this paper, we investigate
the doping variation of the charge and orbital dynamics 
in Y$_{1-x}$Ca$_x$VO$_3$ by measurements of 
the polarized optical conductivity and Raman scattering spectra. 
We discuss 
the effect of the GdFeO$_3$-type orthorhombic lattice distortion 
on the evolution of the spin and orbital ordered state 
with hole doping. 
The results are argued by comparing with those for La$_{1-x}$Sr$_x$VO$_3$ 
with the larger one-electron bandwidth and 
the smaller effect of quenched disorder. 

 The format of this paper is as follows. In Sec. \ref{sec2}, 
we present the experimental procedure.
In Sec. \ref{sec3}, we present the experimental results and 
discuss the doping and temperature variation of 
the charge and orbital dynamics. 
This section is composed of three parts. 
In Sec. \ref{sec3a}, the optical conductivity spectra at 
10K are presented and the doping variation of 
the electronic structure of the ground state is discussed. 
In Sec. \ref{sec3b}, the temperature variation of 
the optical conductivity spectra is presented. 
We discuss the effect of 
the GdFeO$_3$-type orthorhombic lattice distortion in 
the SO and OO. 
In Sec. \ref{sec3c}, we present the Raman scattering spectra and 
discuss the spin and orbital dynamics 
in the doped system. 
The conclusion of this paper is given in Sec. \ref{sec6}.

\section{Experiment \label{sec2}}
\subsection{\label{sec2a}Sample preparation}
Single crystals of Y$_{1-x}$Ca$_x$VO$_3$ were grown 
by the floating zone method. 
As the starting materials, 
we used Y$_2$O$_3$, CaCO$_3$, and V$_2$O$_5$. 
These powders were mixed and 
calcined at 600$^\circ$C and 900$^\circ$C in flowing Ar/H$_2$ 7\% gas. 
The powders were well ground and calcined again 
at 1100$^\circ$C, and then pressed into a rod, 70mm long and 5mm in diameter, 
which was sintered at 1500$^\circ$C in Ar/H$_2$ 7\% flowing. 
The crystal growth was 
performed with use of a halogen-lamp image furnace at the feed speed of 
20mm/h in flowing Ar gas. 
We performed the powder x-ray diffraction 
measurement of the obtained crystal at room temperature and confirmed 
that the single-phase crystals were obtained. 

\subsection{\label{sec2b}Reflectivity measurements}

 The temperature dependence of the reflectivity spectra at nearly normal 
incidence was measured between room temperature and 10K 
in the energy region of 0.01-5eV with linearly polarized light. 
The crystallographic axes of the samples were determined 
by a back-Laue method, and the sample surfaces were mechanically polished 
with alumina powder. 
In the photon energy region of 0.01-0.7eV 
we used a Fourier transform spectrometer (Bruker-IFS66V). 
In the region of 0.5-5eV we used a grating-type monochromator equipped 
with a microscope. 
We carefully mounted samples to ensure the good thermal contact 
on the thermal conduction type microstat. 
In the region of 3-40eV, 
we carried out the measurement at room temperature (RT) with use of 
synchrotron radiation at UV-SOR, Institute for Molecular Science (Okazaki). 
The RT spectra above 5eV were connected smoothly with the ones 
below 5eV to perform the Kramers-Kronig analysis 
and obtain the optical conductivity spectra $\sigma$($\omega$) 
at respective temperatures. 
For the analysis, we assumed constant reflectivity 
below 0.01eV, and also 
used $\omega^{-4}$ extrapolation above 40eV.

\subsection{\label{sec2c}Raman scattering measurements}

 The Raman spectra were measured using a temperature-variable 
cryostat and the triple-grating spectrometer equipped 
with a microscope and liquid-nitrogen-cooled charge coupled device detector. 
The sample surfaces were mechanically polished 
with alumina powder and annealed at 1000$^\circ$C in Ar/H$_2$ 7\% gas flow. 
The 2.410eV (514.5nm) line from an Ar ion laser 
was utilized for the excitation. 
The spot size of the laser 
for the excitation was about 4$\mu$m, 
while the power was reduced to 0.3mW 
to avoid the heating effect. 
We collected the scattered light in backscattering geometry. 
The polarized geometry is described using the conventional notation 
$k_i(e_ie_s)k_s$, 
where $k_i$ and $e_i$ represent the propagation direction and 
the polarization of the incident light, respectively, 
and $k_s$ and $e_s$ those of the scattering light. 
The spectra were measured with the polarization configuration \yxxy, 
where $x$ and $y$ are the (110) and (1-10) crystal direction 
in the orthorhombic ($Pbnm$) setting. 
All the spectra were calibrated by the instrumental sensitivity. 

\section{\label{sec3}Results and Discussion }

\subsection{\label{sec3a}The doping variation of the electronic structure in the ground state }
In this section, we discuss the doping variation of the optical conductivity 
spectra at the ground state. 
The ground state of Y$_{1-x}$Ca$_x$VO$_3$ changes 
from the $G$-type spin and $C$-type orbital ordered phase to 
the $C$-type spin and $G$-type orbital ordered one 
with hole doping beyond $x=0.02$. 
\begin{figure}[htbp!]
  \includegraphics[width=3.375in,keepaspectratio=true]{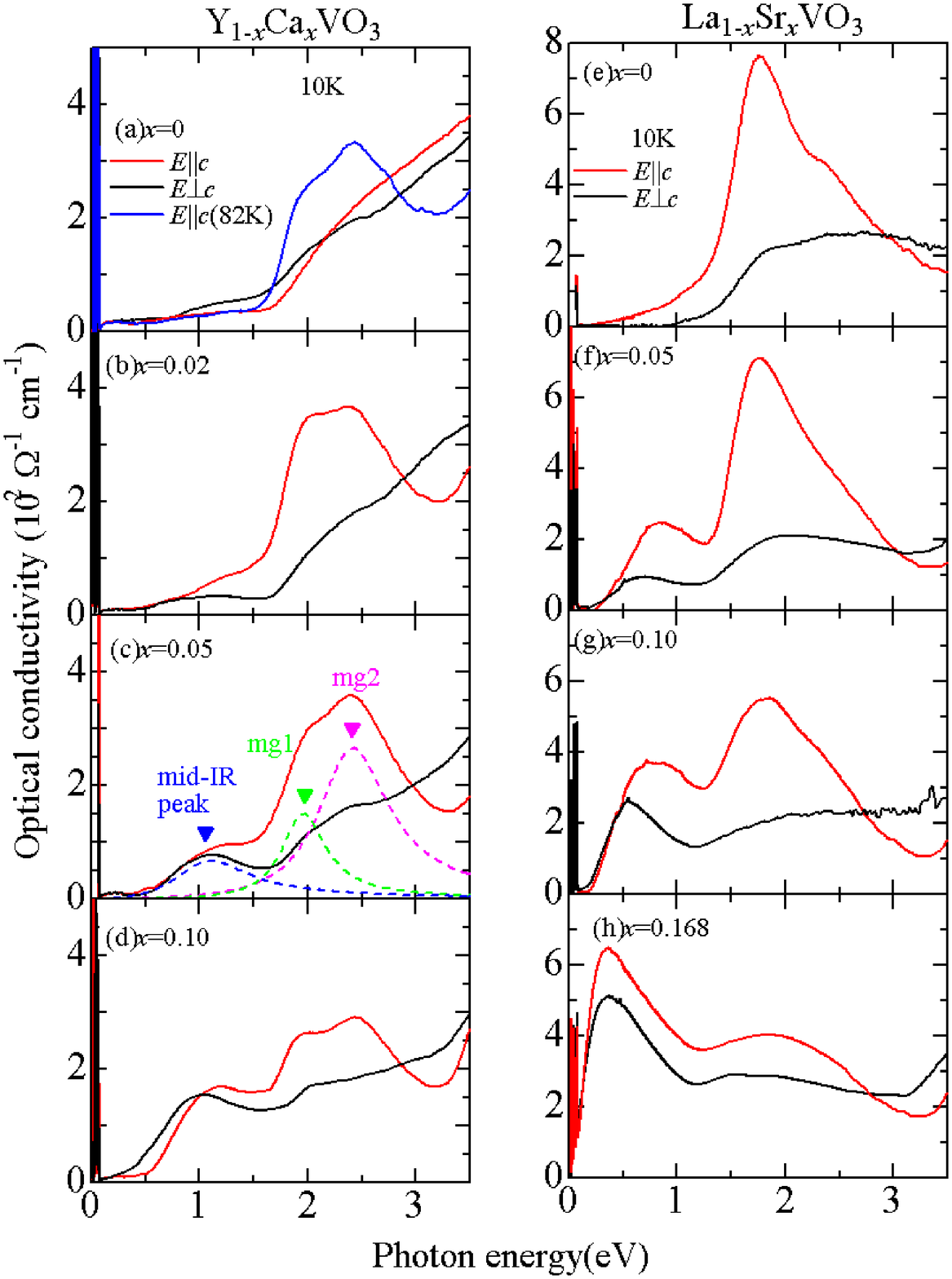}
 \caption{(Color online)Left panel: Optical conductivity spectra at 
10K for \Epara (red line) and \Eperp (black line) 
for Y$_{1-x}$Ca$_x$VO$_3$ with (a)$x=0$, (b)0.02, (c)0.05, and (d)0.10, 
respectively. 
The dashed lines in (c) are 
the Lorentz oscillators (see text) used for the fitting of 
the mid-IR peak and two components of the Mott gap excitation, 
$mg1$ and $mg2$, respectively.
Right panel: Optical conductivity spectra for La$_{1-x}$Sr$_x$VO$_3$ with 
(e)$x=0$, (f)0.05, (g)0.10, and (h)0.168, respectively.}
 \label{fig2}
 \end{figure}
In Fig. 2, we show the polarization dependence of 
the optical conductivity spectra of Y$_{1-x}$Ca$_x$VO$_3$ 
with $x=0$, 0.02, 0.05, 
and 0.10 at 10K and 
reproduce those of La$_{1-x}$Sr$_x$VO$_3$ with $x=0$, 0.05, 0.10, and 0.168 
for comparison.\cite{JF2006} 
For $x=0$ in Y$_{1-x}$Ca$_x$VO$_3$, 
both the \Epara and \Eperp spectra show the clear onset of 
the conductivity at around 1.5 eV, which is assigned to 
the Mott gap excitation in the $3d$ $t_{2g}$ band of V ions 
\cite{Inaba,miya_opt, Fang}, 
and the shape of the spectra is nearly isotropic. 
This behavior is contrasted with the spectral shape for $x=0.02$, 
where the SO and OO are the $C$-type and $G$-type, respectively. 
For $x=0.02$, 
the spectra show a clear peak structure of the Mott-gap excitation 
around 2.5eV for $E\parallel c$, while those for \Eperp appear to 
be nearly identical to those for $x=0$. 
Such an anisotropic feature is similarly observed in the spectra of LaVO$_3$ 
at 10K and YVO$_3$ at $T>T_{SO2}$. 
The recent theoretical calculation has shown that 
the Mott gap excitation strongly depends on 
the spin and orbital exchange interaction and 
that the anisotropic Mott-gap excitation in the $C$-type spin and $G$-type 
orbital ordered phase originates from the one-dimensional 
orbital exchange interaction. \cite{motome, kha_opt} 
In other words, the nearly isotropic spectra of YVO$_3$ ($x$=0) at 10K 
imply that 
the orbital exchange interaction in the 
$G$-type spin ordered state is nearly isotropic in magnitude. 
The present result for YVO$_3$ is not consistent with that of 
the previous paper by Miyasaka {\it et al.}\cite{yvo_opt}, but 
rather in accord with that of the ellipsometry measurement 
by Tsvetkov {\it et al}.\cite{AATsvetkov2004} 
They reported that 
the anisotropic feature of the Mott-gap excitation is reduced 
in the $G$-type spin and $C$-type orbital ordered phase.

For $x=0.05$, in addition to the reminiscence of the Mott-gap excitation 
around 2.5eV, 
a new peak structure shows up around 1eV. 
A similar peak structure is observed in La$_{1-x}$Sr$_x$VO$_3$ 
with $x=0.05$, 0.10, and 0.168, 
as shown in Figs. 2 (f), (g), and (h), and attributed 
to the dynamics of the doped hole. 
It is anticipated from the highly insulating charge transport that 
the doped hole is trapped to form the small polaron 
like state accompanying the lattice relaxation 
with the modification in the spin and orbital sectors.\cite{Inaba, JF2006} 
In this paper, we call the new peak "mid-IR peak" in the following. 
Although the reminiscence of the Mott-gap excitation clearly shows an 
anisotropic feature, 
the mid-IR peak appears nearly isotropic. 
This is contrasted with the case of La$_{1-x}$Sr$_x$VO$_3$, 
where the mid-IR peak 
also shows anisotropy as well as the Mott-gap excitation 
in such a lightly doped region as $x=0.05$. 
For Y$_{1-x}$Ca$_x$VO$_3$ with $x=0.10$, 
where the high-temperature $G$-type orbital ordered phase barely subsists, 
the reminiscence of the Mott-gap excitation still shows anisotropy 
and the intensity of the mid-IR peak becomes larger than that for $x=0.05$. 

 To estimate the spectral weight of the mid-IR peak more quantitatively, 
we fitted spectra with the Lorentz-type oscillators 
\begin{eqnarray}
\sigma(\omega)=\sum_{j=m,mg1, mg2, ct}
\frac{S_j\gamma_j\omega^2\omega_j^2}{(\omega^2-\omega_j^2)^2+\gamma_j^2\omega^2},\nonumber
\end{eqnarray}
and calculated the spectral weight, namely the effective number of electrons 
N$_{eff}$($=2m/\pi e^2N\int_0^{\infty}\sigma(\omega)d\omega$).
Here, $m$, $mg1$, $mg2$, and $ct$ represent the component of the mid-IR peak, 
the two components of Mott-gap excitation, and 
the charge-transfer excitation from O$2p$ to V 3$d$ band, respectively. 
As assigned in the previous work\cite{miya_opt}, 
the lower-lying $mg1$ band is an allowed Mott-gap transition 
along the ferromagnetic chain ($\parallel$ $c$) 
in the $C$-type spin ordered phase, 
while the higher-lying $mg2$ band would be nominally forbidden 
in the case of the perfect spin and orbital polarization 
in the cubic perovskite lattice. 
The latter turns to be visible possibly 
due to the imperfect spin and orbital polarization 
and/or to the orthorhombic lattice distortion. 
\begin{figure}[htbp!]
\includegraphics[width=3.375in,keepaspectratio=true]{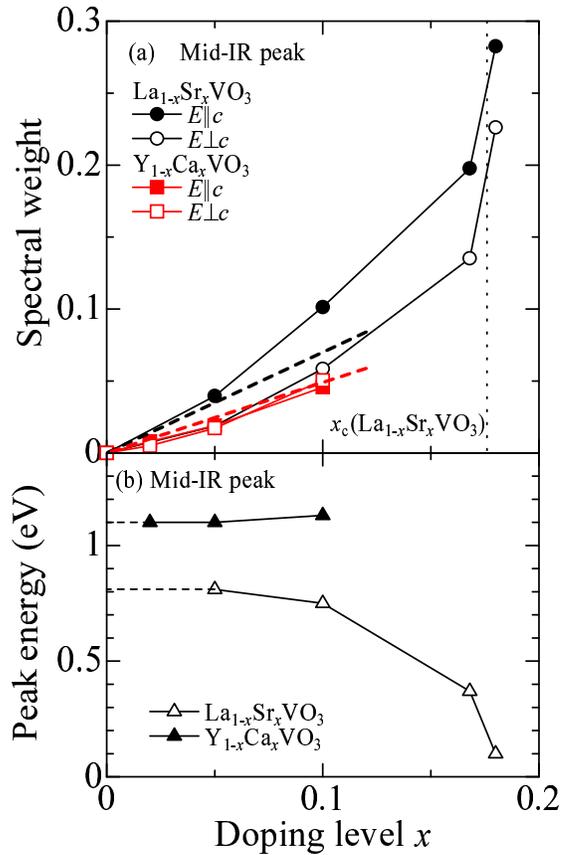}
\caption{(Color online)
 (a) The spectral weight of the mid-IR peak for \Epara 
(closed squares) and \Eperp (open squares) spectra 
in Y$_{1-x}$Ca$_x$VO$_3$ plotted against $x$ 
(See text for the definition of the spectral weight). 
The closed and open circles represent those for \Epara and \Eperp spectra 
in La$_{1-x}$Sr$_x$VO$_3$, respectively. 
The thick dashed lines are the least-square fitting 
of the averaged spectral weight 
N$_{eff}$[av]=(2N$_{eff}$[\Eperp]+N$_{eff}$[\Epara])/3
for $0\le x \le 0.10$ (see text). 
The vertical dotted line indicates the critical doping level 
for the insulator-metal transition ($x_c$=0.176 ) in La$_{1-x}$Sr$_x$VO$_3$. 
(b)The closed and open triangles indicate the energy of 
the mid-IR peak for Y$_{1-x}$Ca$_x$VO$_3$ and La$_{1-x}$Sr$_x$VO$_3$ 
as a function of $x$, respectively.}
 \label{fig3}
 \end{figure}
In Fig. 3 (a), we show the doping variation of 
N$_{eff}$ of the mid-IR peak (N$_{eff}^m$) for \Epara and \Eperp 
in Y$_{1-x}$Ca$_x$VO$_3$ 
as well as that in La$_{1-x}$Sr$_x$VO$_3$. 
The both N$_{eff}^m$ values for \Epara and \Eperp increase monotonously 
with the increase of $x$. 
It should be noted again that 
N$_{eff}^m$ for \Epara is comparable to
that for \Eperp in Y$_{1-x}$Ca$_x$VO$_3$ with $x\le 0.1$, 
in spite of the anisotropic Mott-gap excitation. 
This is contrasted with the case of La$_{1-x}$Sr$_x$VO$_3$, 
where N$_{eff}^m$ in the lightly doped reigon (e.g. $x=0.05$) 
is larger for \Epara than 
for \Eperp as well as that of Mott-gap excitation. 
To see the variation of the kinetic energy of the doped hole, 
we calculated the averaged spectral weight 
N$_{eff}$[av]=(2N$_{eff}$[\Eperp]+N$_{eff}$[\Epara])/3 
at various $x$ (data are not shown to avoid complexity in the figure) 
and performed least-square fitting for $x\le 0.1$, 
the results of which are shown with thick dashed lines in Fig. 3(a). 
The gradient of the least-square fitted line for Y$_{1-x}$Ca$_x$VO$_3$ 
is around $2/3$ of that for La$_{1-x}$Sr$_x$VO$_3$, 
indicating the smaller kinetic energy of 
the doped hole in Y$_{1-x}$Ca$_x$VO$_3$ than in La$_{1-x}$Sr$_x$VO$_3$. 
This is explained in terms of the variation of 
the effective electron correlation 
$U/W$ ($U$ and $W$ representing Coulombic correlation energy 
and the one-electron bandwidth, respectively). 
The larger GdFeO$_3$-type orthorhombic lattice distortion 
in Y$_{1-x}$Ca$_x$VO$_3$ than that in La$_{1-x}$Sr$_x$VO$_3$ results in  
the smaller V-O-V bond angle, leading to the larger effective $U/W$ 
for the V $3d$ $t_{2g}$ band, 
or the suppression of the hole kinetic energy. 
A similar relation of the kinetic energy of the doped hole 
with the effective $U/W$ has also been examined 
for the filling-controlled perovskite titanates 
$R_{1-x}A_x$TiO$_3$ \cite{Katsu1995}. 
In general, the electron-lattice coupling plays a cooperative role 
in localizing the charged carrier with the electron correlation. 
By means of the dynamical mean-field theory, 
Millis {\it et al.} have pointed out that 
the lattice relaxation around the localized electron 
increases with the reduction of the kinetic energy 
via the dimensionless parameter $\lambda=g^2/kt$ 
(here, $g$ is the coupling constant of electron-phonon interaction, 
$k$ the elastic energy of the lattice, 
and $t$ the transfer energy of electron)\cite{Millis1996}. 
In the present system, it is expected that 
the increase in $U/W$ enhances the local lattice relaxation 
accompanying the local modification in the 
spin and orbital sectors around the doped hole. 
Since the dynamics of the doped hole is sensitive to the spin and orbital 
state on the neighboring sites, 
this could explain the 
less anisotropic hole dynamics 
in Y$_{1-x}$Ca$_x$VO$_3$ than in La$_{1-x}$Sr$_x$VO$_3$. 
In addition, 
the enhanced modification in the spin and orbital sectors 
tends to destabilize the long-range SO and OO. 
Thus, 
as well as the quenched disorder in the lattice sector as argued in 
Sec. \ref{sec1}, 
this may also explain 
why the critical doping level required for the melting of the $G$-type OO 
is lower in Y$_{1-x}$Ca$_x$VO$_3$ ($x_o$=0.10) 
than in La$_{1-x}$Sr$_x$VO$_3$ ($x$=$x_c$=0.176).

In Fig. 3 (b) is plotted the $x$ variation of the mid-IR peak energy 
for \Epara spectra ($\omega_{m\parallel}$) 
in Y$_{1-x}$Ca$_x$VO$_3$ compared with that in 
La$_{1-x}$Sr$_x$VO$_3$ 
as a measure of the excitation energy of the doped hole from 
the self-trapped state. 
In both Y$_{1-x}$Ca$_x$VO$_3$ and La$_{1-x}$Sr$_x$VO$_3$, 
the $\omega_{m\parallel}$ is nearly constant for $x\le 0.1$, 
indicating that the doped holes are well localized and 
the interaction among them is negligible. 
In the dilute limit ($x\rightarrow0$)
the $\omega_{m\parallel}$ in Y$_{1-x}$Ca$_x$VO$_3$  
seems to be larger than that in La$_{1-x}$Sr$_x$VO$_3$. 
This is in accord with the anticipation that 
the larger lattice relaxation 
accompanying the larger modification in the spin and orbital sectors would 
trap doped holes at a deeper energy-level state. 

\subsection{\label{sec3b}Temperature variation of the optical conductivity spectra for Y$_{1-x}$Ca$_x$VO$_3$  }

In this section, we present the temperature variation of 
the optical conductivity spectra for Y$_{1-x}$Ca$_x$VO$_3$ 
in comparison with the case for La$_{1-x}$Sr$_x$VO$_3$ 
and discuss the effect of the GdFeO$_3$-type orthorhombic lattice distortion 
on the spin and orbital ordered state. 
Since the optical conductivity spectra for \Eperp show minimal 
temperature variation, we focus on those for \Epara 
in the following discussion. 
\begin{figure}[htbp!]
\includegraphics[width=3.375in,keepaspectratio=true]{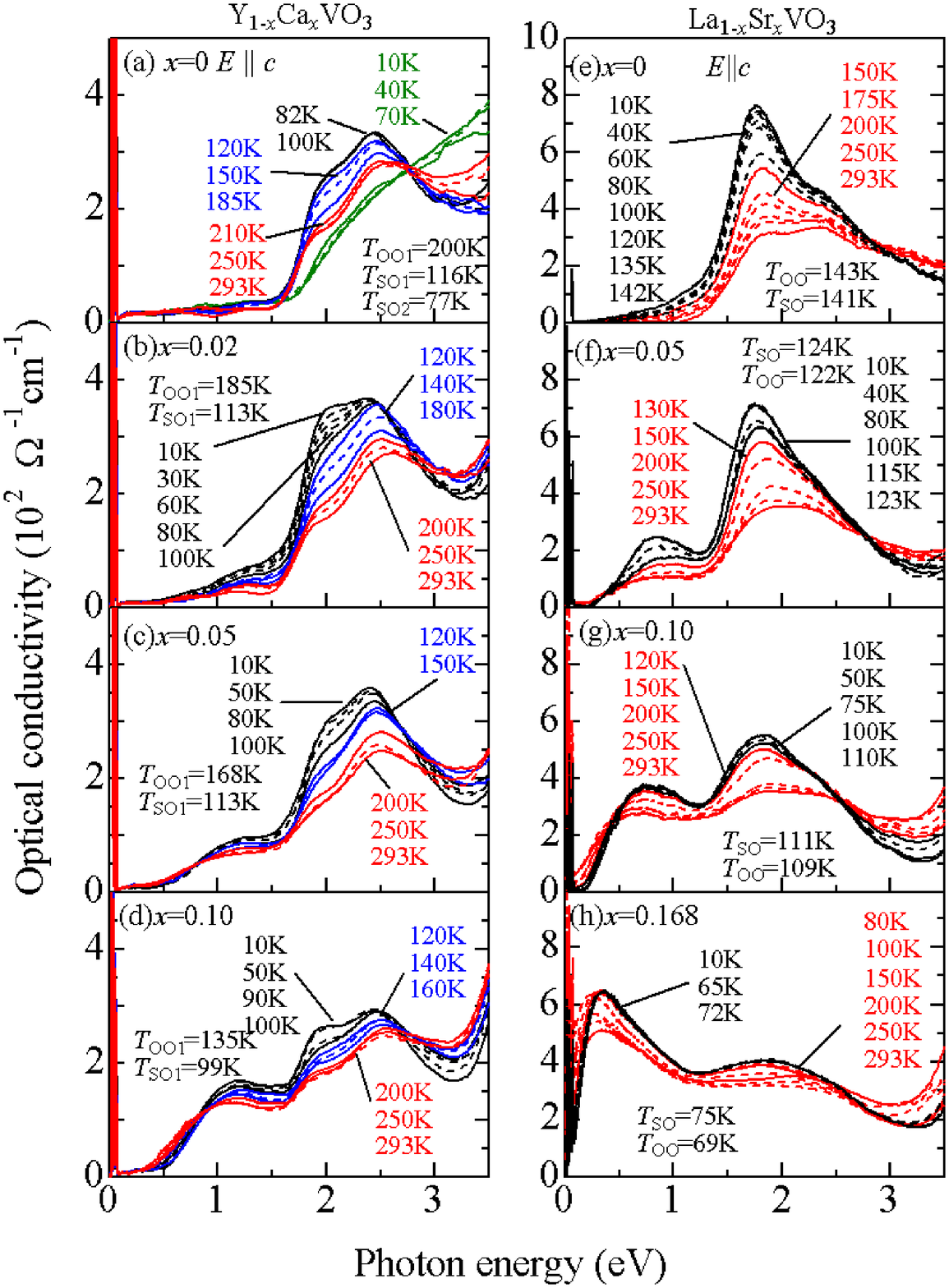}
\caption{
(Color online)
Left panel: The temperature dependence of the optical conductivity spectra 
for Y$_{1-x}$Ca$_x$VO$_3$ with (a)$x=0$, (b)0.02, (c)0.05, and (d)0.10, 
respectively. 
Right panel: The temperature dependence of the optical conductivity spectra 
for \Epara in La$_{1-x}$Sr$_x$VO$_3$ with (e)$x=0$, (f)0.05, (g)0.10, and 
(h)0.168, respectively. }
 \label{fig4}
 \end{figure}
Figures 4 (a)-(d) show the temperature variation of 
the optical conductivity spectra for \Epara in 
Y$_{1-x}$Ca$_x$VO$_3$ with $x=0$, 0.02, 0.05, and 0.10, respectively. 
For $x=0$, 
the spectra show minimal temperature variation 
within the $G$-type spin and $C$-type orbital ordered phase, 
$i.e.$ at $T<T_{SO2}(=T_{OO2})$. 
At $T > T_{SO2}$, as mentioned in Sec. \ref{sec3a}, 
the spectra show a clear peak structure around 2.5eV, 
which reflects the one-dimensional orbital exchange interaction 
and decreases in its magnitude 
with the increase of temperature. 
Such a conspicuous suppression of the intensity of the Mott-gap excitation 
with the increase of temperature is similar to the case for LaVO$_3$, 
as shown in Fig. 4 (e), 
and attributed to the crossover from the one-dimensional to 
three-dimensional exchange interaction 
in the orbital sector.\cite{miya_opt, motome, kha_opt} 
In the doped systems, such as $x=0.02$, 0.05, and 0.10, 
the intensity of the Mott-gap excitation monotonously decreases 
with the increase of temperature and the magnitude of 
the temperature variation also decreases with the increase of $x$. 
A similar behavior is observed in La$_{1-x}$Sr$_x$VO$_3$ and attributed to 
the reduction of the spin and orbital correlation.\cite{JF2006} 

 To estimate the temperature variation of the spectral weight of 
the Mott-gap excitation more quantitatively, 
we calculated N$_{eff}$ of the Mott-gap excitation 
(N$_{eff}^{mg}$) by using the fitting formula 
described in the Sec.\ref{sec3a}. 
\begin{figure}[htbp!]
\includegraphics[width=3.375in,keepaspectratio=true]{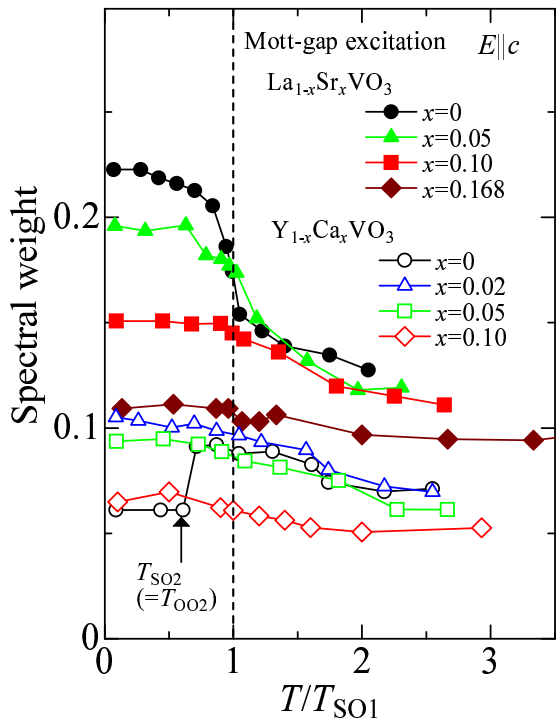}
\caption{
(Color online)
(a) The spectral weight of the Mott-gap excitation N$_{eff}^{mg}$ 
for \Epara spectra 
in La$_{1-x}$Sr$_x$VO$_3$ with $x=0$, 0.05, 0.10, and 0.168 
(closed circles, triangles, squares, and diamonds, 
respectively) and in Y$_{1-x}$Ca$_x$VO$_3$ with $x=0$, 0.02, 0.05, and 0.10 
(open circles, triangles, squares, and diamonds, respectively) 
plotted against the normalized temperature $T/T_{SO1}$. 
}
 \label{fig5}
 \end{figure}
In Fig. 5, 
are plotted the values for Y$_{1-x}$Ca$_x$VO$_3$ and La$_{1-x}$Sr$_x$VO$_3$ 
at various $x$ 
as a function of the normalized temperature ($T/T_{SO1}$).
Here, N$_{eff}^{mg}$ is defined as the sum of the two components of 
the Mott-gap excitation, $mg1$ and $mg2$. 
For LaVO$_3$ ($x$=0), 
N$_{eff}^{mg}$ steeply decreases around $T_{SO1}$ with the increase of 
temperature. 
This behavior is consistent with 
the recent x-ray diffraction study by Y. Ren {\it et al}. \cite{Ren2003}
They reported that the phase transition to the $G$-type OO at $T_{OO1}$ 
in LaVO$_3$ is of the first order. 
With the increase of $x$ in La$_{1-x}$Sr$_x$VO$_3$, 
the steep change of N$_{eff}^{mg}$ around $T_{SO1}$ is extinguished, 
although the temperature variation is still visible at $x=0.168$, 
$i.e.$ on the verge of the insulator-metal transition 
accompanying the orbital order-disorder transition. 

In Y$_{1-x}$Ca$_x$VO$_3$, 
the temperature variation of N$_{eff}^{mg}$ is different from 
that in La$_{1-x}$Sr$_x$VO$_3$. 
For $x$=0, N$_{eff}^{mg}$ shows a clear jump at 
$T_{SO2}$, reflecting the onset of the one-dimensional orbital exchange 
interaction inherent in the $C$-type spin and $G$-type orbital ordered phase. 
At $T>T_{SO2}$, with the increase of temperature, 
N$_{eff}^{mg}$ monotonously decreases 
up to the room temperature as well as that in La$_{1-x}$Sr$_x$VO$_3$. 
In the doped systems Y$_{1-x}$Ca$_x$VO$_3$ with $x=0.02$, 0.05, and 0.10, 
a similar behavior is observed, apart from the 
absence of the $G$-type spin and $C$-type orbital ordered phase 
as the ground state. 
It should be noted that 
N$_{eff}^{mg}$ is less temperature-dependent for 
Y$_{1-x}$Ca$_x$VO$_3$ than for La$_{1-x}$Sr$_x$VO$_3$, 
at the respective doping levels $x$. 
This suggests that 
the increasing GdFeO$_3$-type orthorhombic lattice distortion 
tends to reduce the one-dimensionality of the orbital exchange interaction. 
To see the influence of the GdFeO$_3$-type orthorhombic lattice 
distortion more explicitly, we show 
the optical conductivity spectra for Y$_{1-x}$Ca$_x$VO$_3$ with $x=0.02$ 
and LaVO$_3$ at 10K in Fig. 6. 
\begin{figure}[htbp!]
\includegraphics[width=3.375in,keepaspectratio=true]{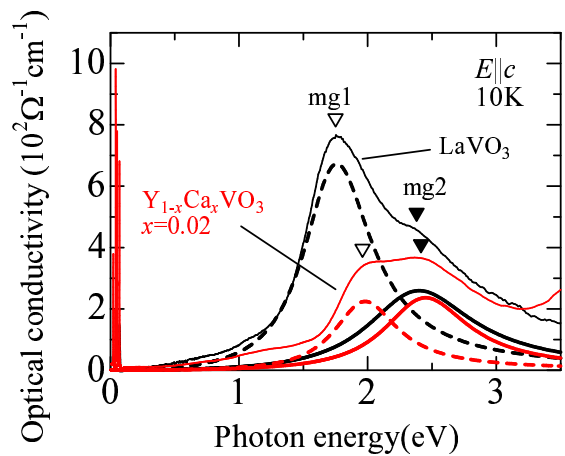}
\caption{(Color online)
Optical conductivity spectra at 10K for \Epara 
in Y$_{1-x}$Ca$_x$VO$_3$ with $x=0.02$ and LaVO$_3$, respectively. 
The thick dashed and solid lines represent the Lorentz oscillators assumed 
for the fitting of the two-components of the Mott-gap excitation, 
$mg1$ and $mg2$, respectively. 
}
 \label{fig6}
 \end{figure}
Since N$_{eff}^{mg}$ for $x=0.02$ is nearly identical to that for $x=0$ 
at $T>T_{SO2}$, 
the reduction of the spin and orbital correlation by the doped hole 
is negligible at such a low doping level. 
Although it is difficult to accurately distinguish 
the spectral weight of $mg1$ from that of $mg2$, 
the former appears to be much smaller in Y$_{1-x}$Ca$_x$VO$_3$ 
with $x=0.02$ than in La$_{1-x}$Sr$_x$VO$_3$ ($x$=0 or 0.05), 
while the latter appear to be nearly comparable. 

On the basis of these results, we discuss the lattice effect 
on the electronic structure. 
The present observation suggests that 
the GdFeO$_3$-type orthorhombic lattice distortion selectively suppresses 
the lower-lying allowed band ($mg$1) relatively to the higher-lying 
(nominally forbidden) one ($mg$2). 
The plausible explanation may be given by the deviation 
of the SO and OO 
from the pure $C$-type and $G$-type, respectively. 
Recently, 
De Raychaudhury {\it et al.} 
performed the $ab$ $initio$ calculation combined with 
the dynamical mean-field theory and 
pointed out that the variation of the crystal field at V sites 
due to the GdFeO$_3$-type orthorhombic lattice distortion 
plays an important role for the OO 
as well as the Jahn-Teller effect.\cite{DeRaychaudhury} 
They showed that 
the GdFeO$_3$-type orthorhombic lattice distortion 
tends to stabilize the $C$-type OO and 
induces the $C$-type character of the OO even in the $G$-type OO 
via the cation-covalency between $R$ and V ions. 
The spin and lattice dynamics coupled with such a modified OO 
is the issue to be discussed in the next section.

\subsection{\label{sec3c}The dynamic spin-orbital phase fluctuation}

In this section, we discuss the doping variation of 
the spin excitation and lattice dynamics coupled with the 
OO in Y$_{1-x}$Ca$_x$VO$_3$. 
As for the undoped systems, 
we recently investigated spin and orbital dynamics 
by measurements of the temperature variation of the 
Raman scattering spectra. \cite{miya_raman} 
\begin{figure}[htbp!]
\includegraphics[width=3.375in,keepaspectratio=true]{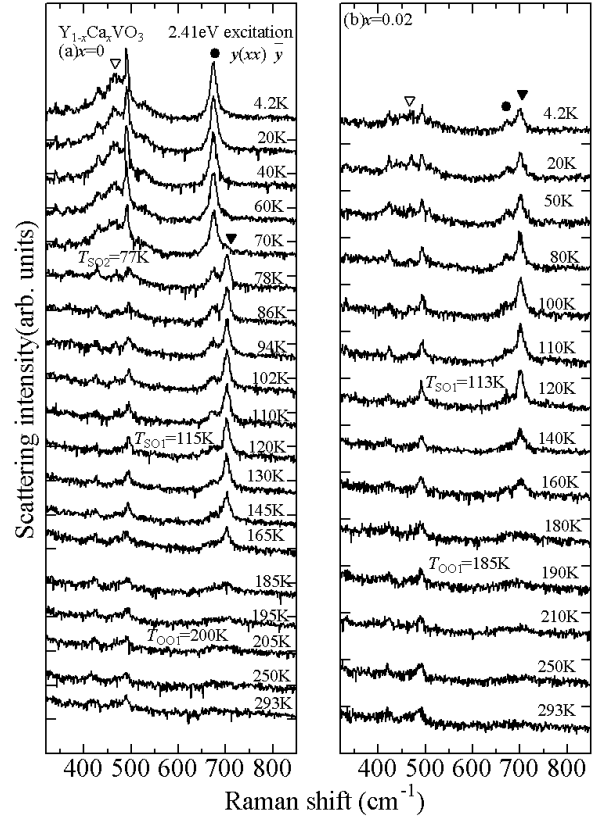}
\caption{(Color online)
Raman scattering spectra for the polarization configuration of \yxxy 
at various temperatures in Y$_{1-x}$Ca$_x$VO$_3$ with 
(a)$x=0$ and (b)$x=0.02$, 
measured with the excitation photon energy of 2.410eV. 
The broad band around 470cm$^{-1}$(open triangle), 
peak structure around 670cm$^{-1}$(closed circle) 
and that around 700cm$^{-1}$(closed triangle) are assigned to 
the two-magnon band, 
the oxygen stretching mode coupled with the $C$-type OO 
and that coupled with the $G$-type OO, respectively. 
}
 \label{fig7}
 \end{figure}
As the starting point of the following discussion, 
we first reproduce in Fig. 7 (a) the temperature variation of 
the Raman scattering spectra for YVO$_3$ with the polarization configuration 
of $y$($xx$)$\bar{y}$, 
which was obtained by utilizing the Ar ion laser 2.410eV (514.5nm) line 
as an excitation light. 
The polarized gyometry is as described in Sec.\ref{sec2c}. 
In the $G$-type spin and $C$-type orbital ordered phase ($T<T_{SO2}$),
the broad two-magnon band and the oxygen stretching mode coupled with 
the $C$-type OO are observed around 470cm$^{-1}$ and 670cm$^{-1}$, 
respectively. 
At $T>T_{SO2}$, the oxygen stretching mode coupled with the $G$-type OO 
appears around 700cm$^{-1}$, but that coupled with the $C$-type OO remains 
to be observed. 
Similarly, the broad two-magnon band is still discerned, 
although its intensity is severely reduced. 
\begin{figure}[htbp!]
\includegraphics[width=3.375in,keepaspectratio=true]{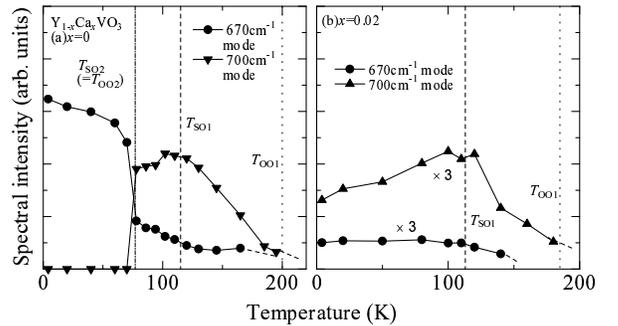}
\caption{(Color online)
Temperature dependence of the spectral intensity of the phonon peaks around 
670cm$^{-1}$ (closed circle) and 700cm$^{-1}$ (closed triangle) 
in Y$_{1-x}$Ca$_x$VO$_3$ with (a)$x=0$ and (b)$x=0.02$, respectively. 
The dash-dotted, dashed, and dotted lines indicate $T_{SO2}$ (=$T_{OO2}$), 
$T_{SO1}$, and $T_{OO1}$, respectively. 
}
 \label{fig8}
 \end{figure}
In Fig. 8 (a), 
the temperature variation of the spectral intensity of each mode 
is reproduced. 
With the increase of temperature, 
the spectral intensity of the 670cm$^{-1}$ mode 
coupled with the $C$-type OO decreases discontinuously at $T_{SO2}$, 
but still remains finite at $T>T_{SO2}$. 
On the other hand, the spectral intensity of the 700cm$^{-1}$ mode 
coupled with the $G$-type OO emerges abruptly above $T_{SO2}$=($T_{OO2}$), 
reaches maximum around 100K, 
and finally decreases monotonously with the increase of temperature. 
These results are interpreted as the subsistence of the $G$-type spin and 
$C$-type orbital component in the nominally $C$-type spin and $G$-type orbital 
ordered phase. \cite{Ulrich, miya_raman} 
In Fig. 7 (b), 
we comparatively show the temperature variation of 
the Raman scattering spectra for $x=0.02$ 
with the polarization configuration of \yxxy measured in the same condition. 
Although the ground state of this system is turned to the 
nominally $C$-type spin and $G$-type orbital ordered phase, 
the broad two-magnon band characteristic of the $G$-type SO 
and the oxygen stretching mode coupled with the $C$-type OO 
are observed around 470cm$^{-1}$ and 670cm$^{-1}$ 
at low temperatures, respectively, 
as in the case of YVO$_3$. 
We also plot the temperature variation of the spectral intensity of 
the two (670cm$^{-1}$ and 700cm$^{-1}$) oxygen stretching modes 
for $x$=0.02 in Fig. 8 (b). 
With increasing temperature, 
the spectral intensity of the 700cm$^{-1}$ mode monotonously 
increases up to around $T_{SO1}$ and decreases at $T > T_{SO1}$, 
while the 670cm$^{-1}$ mode monotonously decreases above $T_{SO1}$.

One possible scenario to explain 
the double peak structure of the oxygen stretching mode 
and the subsistence of the broad two-magnon band 
would be the static phase separation into the two competing phases; 
the phase separation is generally observed in the phase competing systems 
subject to quenched disorder.\cite{Dagotto2005} 
A recent neutron experiment, however, has revealed that 
the magnetic structure of the 
Y$_{1-x}$Ca$_x$VO$_3$ ($x$=0.02) crystal coming from the same batch 
is nearly uniform $C$-type at low temperatures.\cite{Reehuis_rp} 
This excludes the above mentioned phase separation model. 
Thus, a more plausible explanation is the existence of 
the dynamical $G$-type spin and $C$-type orbital correlation. 
This scenario is consistent with both the results of neutron diffraction and 
Raman scattering, 
since the (elastic) diffraction probe 
can hardly capture the rapid dynamical spin fluctuation. 
It should be noted that 
our result is contrasted with the recent x-ray diffraction experiment on 
SmVO$_3$, 
in which the static phase separation between the two competing phases 
is observed down to low temperatures.\cite{MHSage} 
In the present system, however, 
the microscopic phase separation is not favored 
perhaps due to the energy cost at the interface of two competing phases, 
although the inhomogeneity is induced by the well localized doped-hole. 
The unit cell volume 
in the $G$-type spin and $C$-type orbital ordered phase 
is different as much as by 0.16\% 
from that in the $C$-type spin and $G$-type orbital ordered one.\cite{Blake} 
Therefore, the lattice strain at the interface between the two phases 
would cost too much energy to establish the phase-separated state
in the single-domain single crystal.

The $G$-type spin and $C$-type orbital correlation in the present case 
survives even away from the phase boundary. 
In Fig. 9, we show the doping variation of 
the Raman scattering spectra with the polarization configuration of 
\yxxy at 10K. 
\begin{figure}[htbp!]
\includegraphics[width=2.375in,keepaspectratio=true]{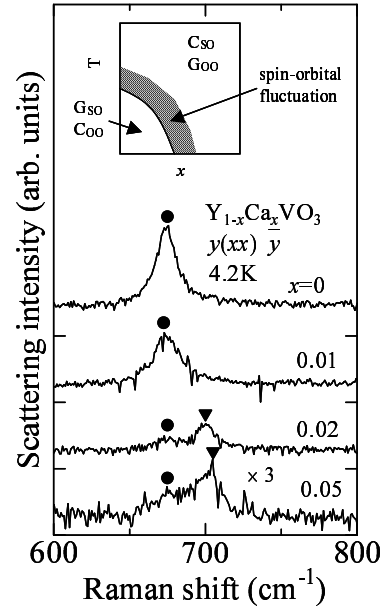}
\caption{(Color online)
Raman scattering spectra at 4.2K in Y$_{1-x}$Ca$_x$VO$_3$ 
with various $x$, measured with the excitation photon energy of 2.410eV. 
The closed circles and triangles indicate the 670cm$^{-1}$ and 700cm$^{-1}$ 
oxygen stretching mode, respectively. 
The inset is the pictorial view of the spin-orbital phase diagram 
in $x-T$ plane in a low temperature region. 
The hatched region indicates the phase, where the dynamical 
$G$-type spin and $C$-type orbital correlation exists.}
\label{fig9}
\end{figure}
With the increase of $x$, 
the spectral intensity of the 670cm$^{-1}$ mode decreases monotonously, 
reflecting the reduction of the correlation of the $C$-type OO, but 
keeps a finite intensity even at $x=0.05$ as well as the 700cm$^{-1}$ one. 
Thus, it is expected that the dynamical $G$-type spin and $C$-type orbital 
correlation remains 
up to a high-doped region at low temperatures, 
as schematically shown in the inset of Fig. 9. 
This is contrasted with the case for La$_{1-x}$Sr$_x$VO$_3$, 
in which the ground state 
remains to be the pure $C$-type spin and $G$-type orbital ordered state 
up to the insulator-metal transition point ($x_c=0.176$).

\section{\label{sec6}conclusion }

In conclusion, 
we have investigated the spin and orbital state in Y$_{1-x}$Ca$_x$VO$_3$ 
by measurements of the optical conductivity and Raman scattering spectra 
with the focus on the hole doping effect on the two competing 
spin and orbital ordered phases. 
We compared the results with those for 
another canonical system La$_{1-x}$Sr$_x$VO$_3$ 
which shows less orthorhombic lattice distortion and 
larger one-electron bandwidth. 
The doped hole forms the self-trapped small polaron like state 
with the lattice relaxation  
accompanying the modification in the spin and orbital sectors. 
The distortion in the spin, orbital and lattice sectors 
appears to be enhanced by the decrease in the one-electron bandwidth, 
{\it i.e.} in going from La$_{1-x}$Sr$_x$VO$_3$ to Y$_{1-x}$Ca$_x$VO$_3$. 
This explains the observed feature that 
the hole dynamics in Y$_{1-x}$Ca$_x$VO$_3$ 
is less sensitive to the evolution of 
the spin and orbital ordering and is nearly isotropic 
even in the lightly doped region. 
This is contrasted with the anisotropic hole dynamics observed 
in the lightly doped region of La$_{1-x}$Sr$_x$VO$_3$. 
The modest temperature variation and small spectral weight of the 
lower-lying allowed Mott-gap excitation suggests that 
the OO deviates from the pure $G$-type. 
At low temperatures, 
the broad two-magnon band and the oxygen stretching mode inherent in 
the $G$-type spin and $C$-type orbital ordered phase 
are persistently observed in the Raman scattering spectra 
for the doping induced phase ($x\ge 0.02$) of the nominally $C$-type SO 
and $G$-type OO, indicating the subsistence of 
the dynamical $G$-type spin and $C$-type orbital correlation.

\section{\label{sec7}acknowledgement }
We thank B. Keimer, C. Ulrich, and M. Reehuis for 
discussion. One of the authors (J. F) was supported by 
the Japan Society for the Promotion of Science for 
Young Scientists. 
This work was in part supported by 
the Grant-In-Aid for Scientific Research (Grants No. 15104006, No. 17340104, 
and 16076205) from MEXT of Japan.


\begin{thebibliography}{99}
\bibitem{yvo_opt}
In the previous paper\cite{miya_opt}, 
the optical conductivity spectra of YVO$_3$ show 
the anisotropic Mott-gap excitation even at $T<T_{SO2}$. 
This discrepancy 
perhaps results from the wrong setting of the sample temperature in 
the previous work. 
A crystal of YVO$_3$ tends to be cracked 
during the measurements in warming or cooling process especially 
around $T=T_{SO2}$, which degrades the thermal conduction 
between the sample surfaces and the cold-finger of the cryostat. 
Thus, we carefully mounted samples and checked the reproducibility 
of the data 
by measuring the spectra several times both in warming and cooling 
process. 
It is likely that the sample surface was not enough cooled down below 
$T_{SO2}$ in the previous work Ref \cite{miya_opt}.
\end{thebibliography}
\end{document}